\documentclass[runningheads,a4paper]{llncs}
\usepackage{algorithm}
\usepackage{algorithmic}
\usepackage{pdflscape}
\usepackage{authblk}
\usepackage{multirow,longtable}
\usepackage{amssymb}
\usepackage{color}
\usepackage{url}
\usepackage[utf8]{inputenc}
\usepackage[T1]{fontenc}
\usepackage{cancel}
\usepackage{tikz}
\usepackage{amsmath,mathtools}
\usepackage{stmaryrd}
\usepackage{caption}
\usepackage{subfig}
\usepackage{rotating}
\usepackage{float}
\usepackage{blindtext}
\usetikzlibrary{shapes.geometric, arrows}
\usepackage{cite}

\begin{document}

\title{Improved Core Genes Prediction for Constructing well-supported Phylogenetic Trees in large sets of Plant Species}
\author{Bassam AlKindy\inst{1,2}, Huda Al-Nayyef\inst{1,2},Christophe Guyeux\inst{1}, Jean-Fran\c{c}ois Couchot\inst{1}, Michel Salomon\inst{1}, Jacques M. Bahi\inst{1}}
\institute{$^1$~FEMTO-ST Institute, UMR 6174 CNRS, Department of Computer Science\\
for Complex Systems (DISC), University of Franche-Comt\'{e}, France,\\
$^2$~Department of Computer Science, University of Mustansiriyah, Baghdad, Iraq,\\
\email{\{bassam.al-kindy, huda.al-nayyef, christophe.guyeux, jean-francois.couchot,	michel.salomon, jacques.bahi\}@univ-fcomte.fr}}
\newcommand{\keywords}[1]{\par\addvspace\baselineskip
\noindent\keywordname\enspace\ignorespaces#1}

\maketitle
\begin{abstract}
The way to infer well-supported phylogenetic trees that precisely reflect the evolutionary process is a challenging task that 
completely depends on the way the related core genes have been found. 
In previous computational biology studies, many similarity based algorithms, mainly dependent on 
calculating sequence alignment matrices, have been proposed to find them.
In these kinds of approaches, a significantly high  similarity score between two coding sequences extracted from a given annotation tool means that one has the same genes.
In a previous work article, we presented a quality test approach (QTA) that improves the core genes quality 
by combining two annotation tools (namely NCBI, a partially human-curated database, and DOGMA, an efficient annotation algorithm for chloroplasts). 
This method takes the advantages from both sequence similarity and gene features to guarantee that the core genome contains correct and well-clustered coding sequences (\emph{i.e.}, genes). 
We then show in this article how useful are such well-defined core genes for biomolecular phylogenetic reconstructions, by investigating various subsets of core genes at various family or genus levels, leading to subtrees with strong bootstraps that are finally merged in a well-supported supertree.

\begin{keywords}
Quality test,
Phylogenetic tree,
Bootstrap,
RAxML,
Core genome,
Core genes,
Supertree.
\end{keywords}

\end{abstract}

\section{Introduction}\label{sec:intro}
Given a collection of genomes, it is possible to define their core genes as the common genes that are shared among all the species, while pan genome is all the genes that are present in at least one genome (\emph{all} the species have each core gene, while a pan gene is in \emph{at least one} genome).

The key idea behind identifying core and pan genes is to understand the evolutionary process among a given set of species: the common part (that is, the core genome) is of importance when inferring the phylogenetic relationship, while accessory genes of pan genome explain in some extend each species specificity. We introduced in a previous study~\cite{Alkindy2014} two methods for discovering core and pan genes of chloroplastic genomes using both sequence similarity and alignment based approaches. Later, we presented in another study~\cite{Alkindy_BIBM2014} the quality test approach as a method to find the core genes for chloroplast species. This article is an extended version of~\cite{Alkindy_BIBM2014,Alkindy2014} focusing on how the quality core genes will affect the phylogenetic inference, and also a performance analysis in terms of execution time and memory consumption. 

Chloroplasts is one of many types of organelles in the plant cell. They are considered to have originated from cyanobacteria through endosymbiosis, when an eukaryotic cell engulfed a photosynthesizing cyanobacterium, which remained and became a permanent resident in the cell. The term of chloroplast comes from the combination of plastid and chloro, meaning that it is an organelle found in plant cell that contains the chlorophyll. Chloroplast has the ability to take water, light energy, and carbon dioxide ($CO_2$) to convert it in chemical energy by using carbon-fixation cycle~\cite{chaffey2003alberts}~(also called \textit{Calven Cycle}, the whole process being called photosynthesis).
 This key role can explain why chloroplasts are at the basis of most trophic chains and thus responsible for evolution and speciation. Moreover, as photosynthetic organisms release atmospheric oxygen when converting light energy into chemical energy and simultaneously produce organic molecules from carbon dioxide, they originated the breathable air and represent a mid to long term carbon storage medium. Consequently, exploring the evolutionary history of chloroplasts is of great interest and therefore further phylogenetic studies are needed.   

A key idea in phylogenetic classification is that a given DNA mutation shared by at least two taxa has a larger probability to be inherited from a common ancestor than to have occurred independently. Thus shared changes in genomes allow to build relationships between species. Homologous genes are genes derived from a single ancestral one. These genes are divided in two types, namely paralogous and orthologous. Paralogous genes arise from ancestral gene duplication while the orthologous genes are products of speciation.
In the case of chloroplasts, an important category of genomes changes is the loss of functional genes, either because they become ineffective or due to a transfer to the nucleus. Thereby a small number of genes lost  among species may indicate that these species are close to  each other and belong to a similar lineage, while a  large lost  means distant lineages. Phylogenies of photosynthetic plants are important to assess the origin of chloroplasts and the modes of gene loss among lineages. These phylogenies are usually done using a few chloroplastic genes, some of them being not conserved in all the taxa. This is why selecting core genes may be of interest for a new investigation of photosynthetic plants phylogeny. 
 
To determine the core of chloroplast genomes for a given set of photosynthetic organisms, bioinformatics investigations using sequence annotation and comparison tools are required, and therefore various choices are possible. The purpose of our research work is precisely to study the impact of these choices on the obtained results. A state of the art for core genome discovery studies is detailed in Section~\ref{sec:stateofart}, whereas a general presentation of the approaches we propose is provided in Section~\ref{sec:general}. To make this paper standalone, a closer examination of the approaches is given in Section~\ref{sec:extraction}, where we will present coding sequences clustering method based on sequence similarity, and quality test method based on quality genes. Information regarding computation time and memory usage is provided in Section~\ref{sec:implem}, while an application example in the field of phylogeny is illustrated in Section~\ref{sec:phylogeny}. This research work ends with a conclusion section summarizing our investigations and giving suggestions for future work. 

\section{State of the art}\label{sec:stateofart}
An early study of finding the common genes in chloroplasts was realized in 1998 by \emph{Stoebe et al.}~\cite{stoebe1998distribution}. They established the distribution of 190 identified genes and 66 hypothetical protein-coding genes (\emph{ysf}) in all nine photosynthetic algal plastid genomes available (excluding non photosynthetic \emph{Astasia tonga}) from the last update of plastid genes nomenclature and distribution. The distribution reveals a set of approximately 50 core protein-coding genes retained in all taxa. \emph{Grzebyk et al.}~\cite{grzebyk2003mesozoic}, for their part, have studied in 2003 the core genes among 24 chloroplastic sequences extracted from public databases, 10 of them being algae plastid genomes. They broadly clustered the 50 genes from \emph{Stoebe et al.} into three major functional domains: (1) genes encoded for ATP synthesis (\emph{atp} genes); (2) genes encoded for photosynthetic processes (\emph{psa} and \emph{psb} genes); and (3) housekeeping genes that include the plastid ribosomal proteins (\emph{rpl} and \emph{rps} genes). The study shows that all plastid genomes were rich in housekeeping genes with one \emph{rbcLg}  gene involved in photosynthesis.     
In 2014, \emph{De Chiara et al.}~\cite{de2014genome} aligned all of the 97 sequenced genomes to a reference, the complete genome of the \emph{Haemophilus influenza} strain 86-028NP, using the Nucmer alignment program~\cite{kurtz2004versatile}. They generated a list of polymorphic sites with these alignments. This list was then filtered to include only the polymorphic sites in the core genome of NTHi, \emph{i.e.}, the regions of the reference strain that could be aligned against all other strains, yielding a set of 149,214 SNPs. A clustering algorithm has been finally used on these SNPs to achieve core genes extraction. Remark that most of these studies used only a low amount of plant genomes to extract the core genome.

\section{An overview of the Pipeline}\label{sec:general}


In previous work~\cite{Alkindy2014}, an annotation based method has been presented in a pipeline for core genomes discovery. It is based on an Intersection Core Matrix 
(ICM) using gene features like gene names. The produced core tree
has then been compared to the phylogenetic one. However, working with gene features alone does not lead to accurate core genomes. This is because of two reasons: first, gene name does not necessary point to the same sequence among different genomes. Second, gene features in the absence of gene sequences cannot provide information such as starting and ending codons, mutation rate, proteins, and so on.  
Such limitations in core genomes confidence is the main reason explaining why we are investigating a new direction.
This new proposal consists of a pre-processing step using a Needleman-Wunch global
alignment, before taking into account gene features, see 
\begin{figure}[!ht]  
  \centering
    \includegraphics[width=0.4\textwidth]{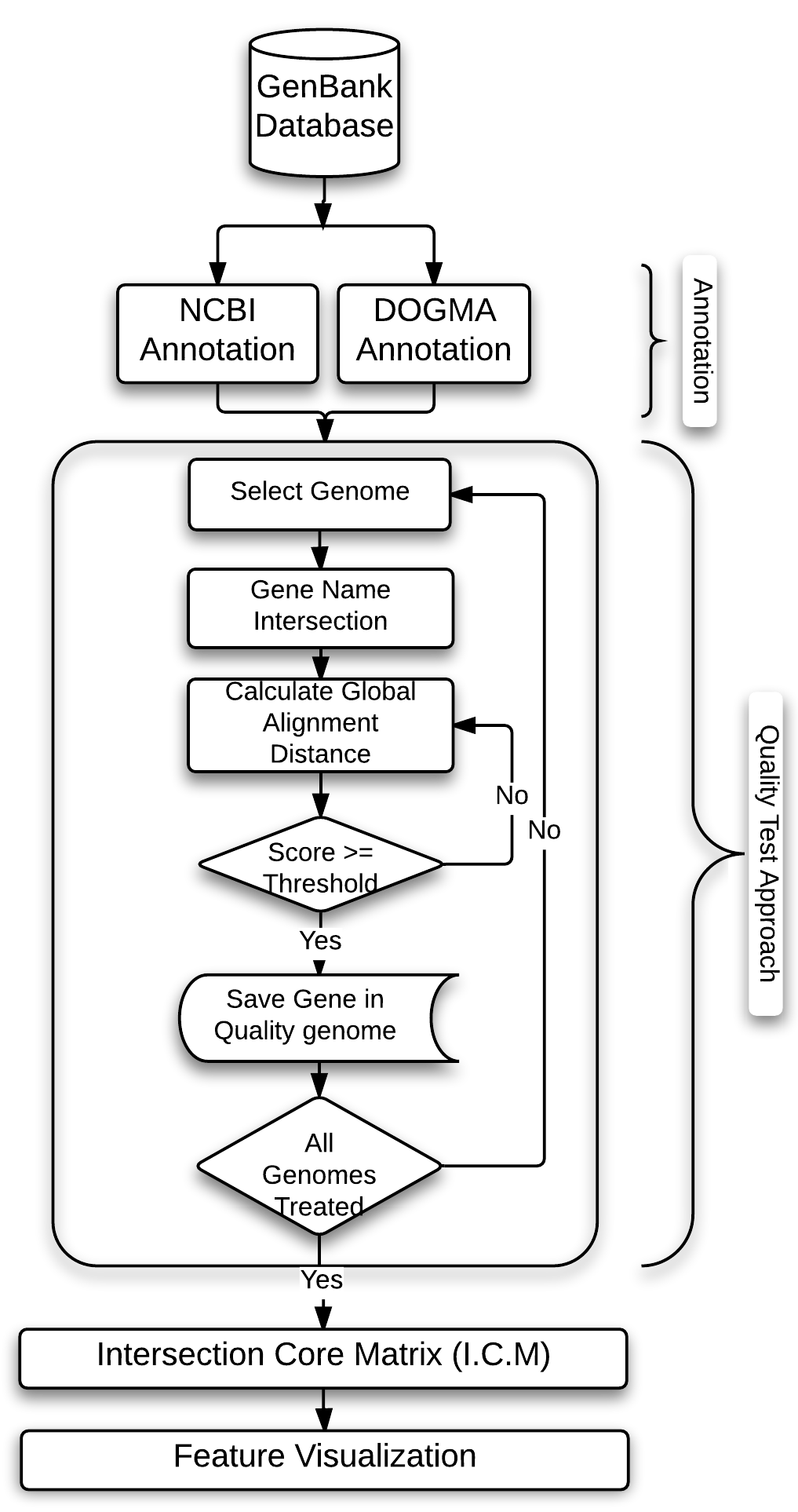}
\caption{An overview of the pipeline~\cite{Alkindy_BIBM2014}.}\label{Fig1}
\end{figure}
Figure~\ref{Fig1}. 

As a starting point, annotations (DNA coding sequences with gene
names and locations) must be provided on the set
of chloroplastic genomes under considerations. Obviously, this annotation stage must be of quality if we want
to obtain acceptable core and pan genomes.
Such a stage necessitates a DNA sequences database 
like NCBI's GenBank
, the   European \textit{EMBL} database~\cite{apweiler1985swiss}, or the Japanese  \textit{DDBJ} one~\cite{sugawara2008ddbj}.
The annotations can be directly downloaded
from these websites, however it is  
preferable to launch an \emph{ad hoc} 
annotation tool on complete downloaded  genomes, like the DOGMA one~\cite{RDogma}. 

Using such annotated genomes, we will employ two general approaches detailed in previous studies for extracting the core genome, which represent the second stage of the pipeline: the first approach involves similarity values that are computed on predicted coding sequences, while the second one takes benefits from all the information provided during the annotation stage.
 
Instead of considering only gene sequences taken from NCBI or DOGMA, we consider to use an improved quality test process provided in~\cite{Alkindy_BIBM2014} in this new proposal. It works with gene names and sequences, to produce what we call ``quality genes''. Remark that such a simple general idea is not so easy to realize, and that it is not sufficient to only consider gene names returned by such tools.
Providing good annotations is  an important stage for extracting gene features. Indeed, gene features here could be considered as: gene names, gene sequences, protein  sequences, and so on. 
We will subsequently propose  methods that use gene names and sequences for extracting core  genes and  producing  chloroplast evolutionary tree.

An extra step named \textit{feature visualization}~\cite{Alkindy2014}, will be added  as a final stage of our pipeline. The construction of core tree and/or phylogenetic tree is done by taking the advantage
of information produced during the core and pan genomes search. This feature visualization stage will then be used to encompass phylogenetic tree construction using core genes, genes content evolution illustrated by core trees, functionality investigations, and so on. 
For illustration purposes, we have considered 
99~genomes of chloroplasts downloaded from GenBank database.

\noindent These genomes cover  eleven types  of chloroplast families (see~\cite{Alkindy2014}). 
Furthermore, two kinds of annotations will be considered in this work, namely the
ones provided by NCBI on the one hand, and the ones by DOGMA on the other hand.

\section{Core genes extraction}\label{sec:extraction}
In this section, we consider the gene prediction approach based on sequence similarity presented in~\cite{Alkindy2014,Alkindy_BIBM2014}. This method starts with genomes annotated, either from NCBI or DOGMA, and uses a distance on genes coding sequences \linebreak
$d:N=\{A,T,C,G\}^{\ast}\times\{A,T,C,G\}^{\ast}\rightarrow[0,1]$,
where $A^{\ast}$ is the set of words on alphabet $N$, 
to group similar alleles  in a same cluster. 

Let us now present the proposed quality test improvement.
The inputs are genomes annotated twice, by NCBI and DOGMA respectively. To extract the common genes, a post-treatment of 
these annotations must first be achieved. On the NCBI side, due to the large variety of annotation origins (being produced either by human or by various automatic tools), 
we have to compute an edit similarity distance on gene names. 
The same name is then set to sequences whose names are close according to this edit distance.
This stage is not required in the DOGMA side, as names are provided by an unique algorithm. However, DOGMA investigates the six reading frames when extracting coding sequences~\cite{RDogma}, and it sometimes produces various fragments for one given gene. So a gene whose name is present at least twice in the file is either a duplicated gene or a fragmented one. 
Obviously, these issues must be fixed and ``fragmented'' genes have to be defragmented before the DNA similarity computation stage (such  defragmentation has normally already been realized on NCBI website). As the orientation of each gene fragment is given in output file, this defragmentation consists in concatenating all the possible permutations, and only keeping the permutation with the best similarity score in comparison with other sequences having the same gene name: this score has to be larger than a given predefined threshold. 

The risk is now to merge genes that are different but whose names are similar (for instance, ND4 and ND4L are two different mitochondrial genes, but with similar names). To fix such a flaw,
the sequence similarity, for intersected genes in a genome, is compared too in a second stage (with a Needleman-Wunsch global alignment) after selecting a genome accession number, and the genes correspondence
is simply ignored if this similarity is below a predefined threshold. We call this operation, which will result in a set of quality genes, a \textit{quality test}. These genes will then constitute the quality genomes. A list of generated quality genomes based on specific threshold is then produced. It is used to construct the intersection core matrix, which will generate the core genes, core tree, and phylogenetic tree after choosing an appropriate outgroup. 
In this work, to improve the confidence put in the core genes, we have discarded the paralogous genes.

\section{Implementation}\label{sec:implem}
All  algorithms have  been implemented using  Python language version 2.7, on a personal computer running Ubuntu~12.04 32~bits with 6~GByte memory,
and a quad-core Intel Core~i5~processor with an operating frequency of 2.5~GHz. 

\subsection{Construction of quality genomes}


The first step in producing 
annotated genomes is to find the set of common genes, that is, genes sharing similar names and sequences, by using various annotation tools and following the method described previously. Figure~\ref{subfig-1:ncbi_vs_dogma} presents the original amount of genes based on NCBI and DOGMA annotations. Two quality test routines then take place to produce ``quality genomes'' by: (1) selecting all common genes based on gene names and (2) checking the similarity of sequences, which must be larger than or equal a predefined threshold (see Figure~\ref{subfig-1:ncbi_vs_dogma}). Note that predefined threshold is not used to determine the ortholog genes, it is used to ensure that core genes from NCBI and DOGMA annotations are identical. We also calculate the correlation function to see with whom the common genes have good relation (\emph{e.g.} with NCBI or Dogma)? We found that the correlation value based on the number of genes produced by two annotation algorithms is 0.57. The correlation value based on the number of genes between the produced quality genomes and NCBI genomes is 0.6731, and 0.9664 between produced quality genomes and Dogma genomes. Note that gene differences between such annotation tools can affect the final core genome, if the naming and the functionality of these genes are well defined.  

\subsection{Core and pan genomes}
\begin{figure}[!ht]
\begin{center}
    \subfloat[Amount of genes based on NCBI and DOGMA w.r.t quality common genes. DOGMA gives the larger number of genes.\label{subfig-1:ncbi_vs_dogma}]{%
    \includegraphics[width=0.65\textwidth]{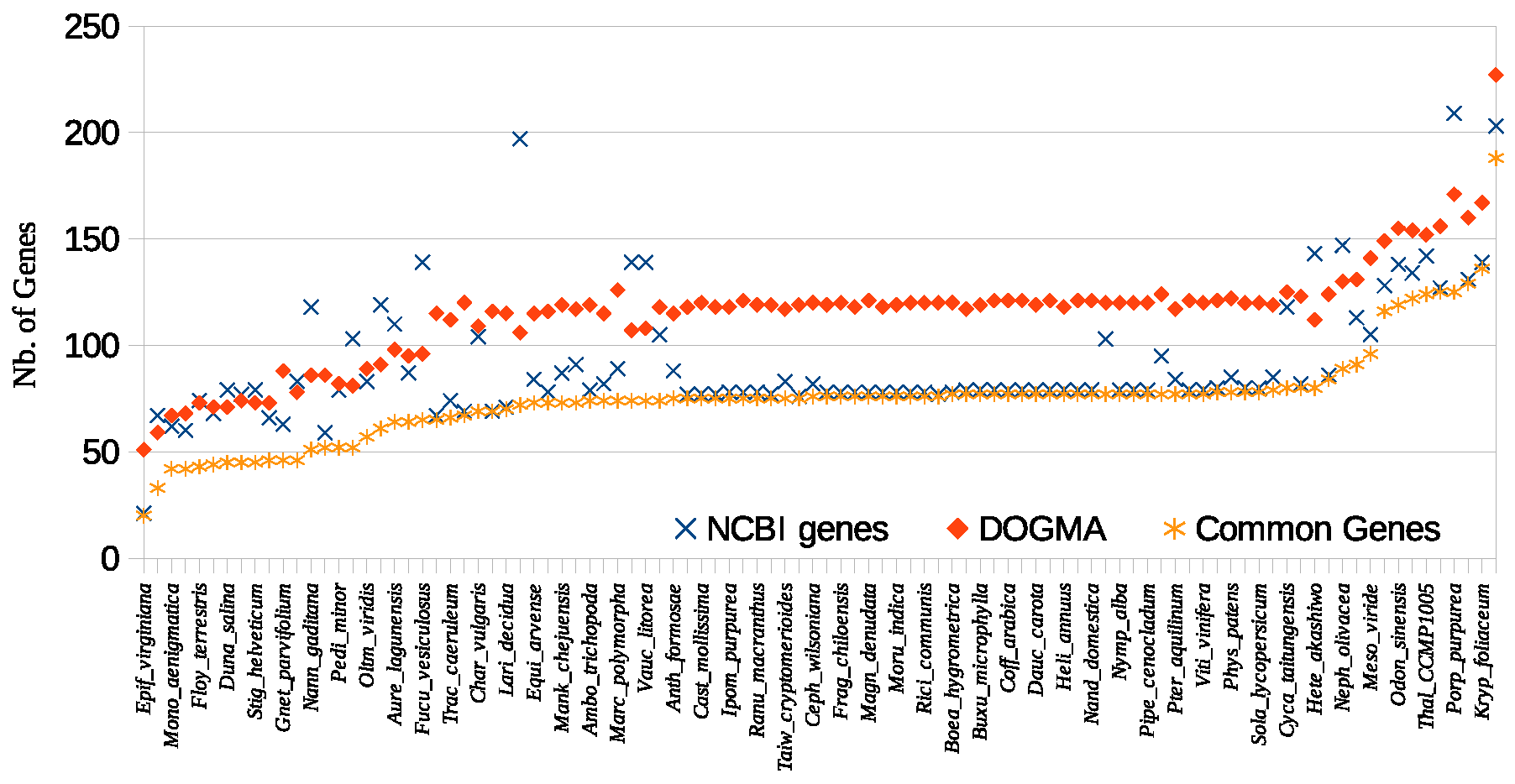} 
    }
    \vfill
    \subfloat[Core genomes sizes w.r.t. threshold. A maximal number of core genes does not mean a good core genomes: we are looking for genes meeting biological requirements.\label{subfig-1:core}]{%
      \includegraphics[width=0.65\textwidth]{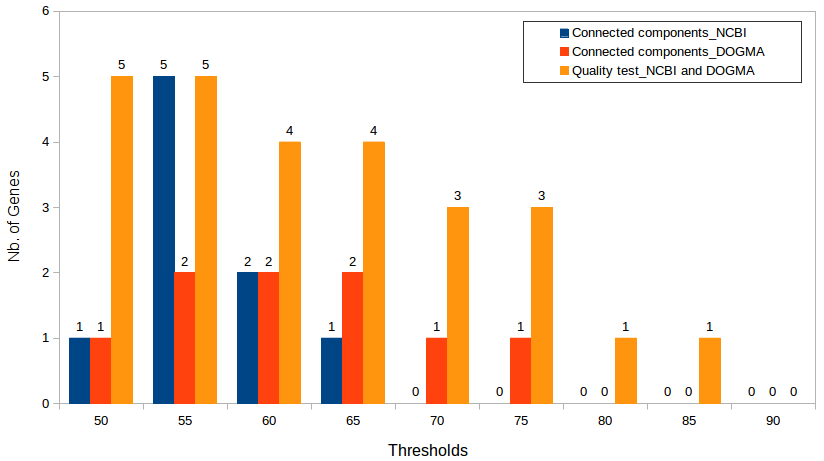} 
    }
    \caption{ (a) Genes coverage for a threshold of 60\% and (b) core genomes sizes.}
    \label{fig:coverage}
\end{center}
\end{figure}
Figure~\ref{subfig-1:core} represents the  amount of genes in  the computed core genome of 98 species. In this figure, two methods are used and compared using the same sample of genomes: in the first one, the gene prediction approach presented in~\cite{Alkindy2014,Alkindy_BIBM2014} has been used on genomes annotated by NCBI, while on the other one the quality test approach~\cite{Alkindy_BIBM2014} has been applied on genomes annotated by DOGMA. Different thresholds have been examined for both approaches. The amount of final core genes within the two approaches is low, as the species considered here are highly divergent. However even in that particular situation, it is obvious that the quality test approach outperforms the other 
one at each tested threshold.
Compare to Coregenes~\cite{zafar2002coregenes}, our approach allows 
to deal with a large class of genomes (98 species) 
whereas this tool is limited 
to six genomes. 
 As stated previously, the main goal is to  find the largest number of core genes compatible
 with biological background related to chloroplasts. 
In the quality approach case, one genome (\emph{Micromonas pusilla}, with accession number NC\_012568.1) has been discarded from the 
sample, as we observed that this genome always has the minimum number of common genes with its correspondents. That can be 
explained by two reasons: (1) either it consists of non-functional genes, or (2) the diversity value is too high. 
With quality approach, an absence of genes in rooted core genome means that we have two or more sub-trees of organisms completely 
divergent among each other.
Unfortunately, for the first approach with NCBI annotation, the core genes within NCBI cores tree did not provide 
a distribution of genomes that are biologically supported. More precisely, \textit{Micromonas pusilla} (accession number NC\_012568.1) 
is the only genome that totally destroys the final core genome with NCBI
annotations, for both gene features and gene quality methods.
Conversely, in the case of DOGMA annotation, the  distribution of genomes is 
biologically relevant\footnote{Core tree is available on~\url{http://members.femto-st.fr/christophe-guyeux/en/chloroplasts}.} 
. 

\subsection{Execution time and memory usage}
\begin{figure}[!ht]
    \subfloat[Time needed to execute each method.
\label{subfig-1:time}]{%
    \includegraphics[width=0.48\textwidth]{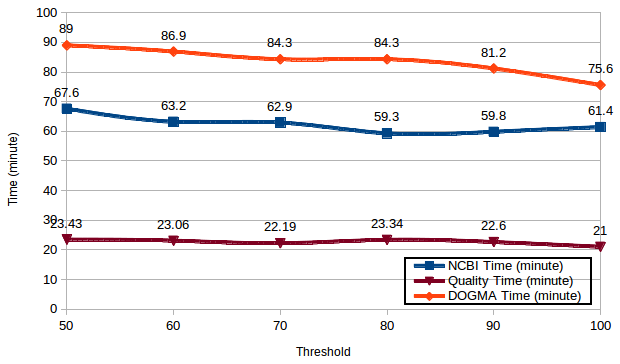}
    }
    \hfill
    \subfloat[Memory usage (MB unit) (sizes usually available on  personal computers).\label{subfig-2:memory}]{%
      \includegraphics[width=0.48\textwidth]{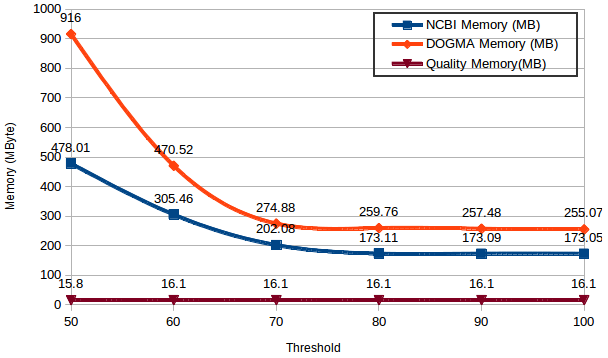}
    }
    \caption{Execution time and memory usage w.r.t. threshold.}
    \label{fig:time_mem}
\end{figure}

In computational biology, time and memory consumptions are two important  factors due to high throughput operations among gene sequences. Figure~\ref{fig:time_mem} shows the amount of time and memory needed to extract core genes using the two approaches: in the first one, 
building the connected components depends on the construction of a distance matrix by considering the similarity scores from the global alignment tool, which takes a long time in the case of NCBI and DOGMA genomes. Calculation time is different for DOGMA and NCBI due to the size of genomes and the amount of gene sequences that need to be compared: NCBI genomes have 8,992 genes, instead of 11,242 in DOGMA genomes. Figure~\ref{subfig-1:time} presents the execution time needed for each method with respect to thresholds in range $[50-100]$. But the DOGMA one requires more computational time (in minutes) for sequence comparisons, while gene quality method needs a low execution time to compare quality genes. However, once the ``quality genomes'' have been constructed, this method takes only 1.29~minutes to extract core genes.

The second important factor is the amount of memory used by each methodology, this one is highlighted by  Figure~\ref{subfig-2:memory}. The low values show that the gene quality method based on gene sequence comparisons presents the
most reasonable  memory  usage (when constructing quality genomes). It also depends on the size of genomes. Determining which method to choose depends on the user preferences: if we search for a fast and semi-accurate method, then the second approach should be chosen. Otherwise, if an accurate but relatively slow approach is desired, then the first method with DOGMA annotations should be preferred.

\section{Phylogenetic study illustration}\label{sec:phylogeny}

To show the relevance of the obtained quality core genes, we will use all or a proportion of them to build a phylogenetic tree. Indeed, thanks to our approach we can precisely identify the common genes of a group of species and thus use the corresponding core genome to deduce their phylogenetic relation. The objective is to find the most well-supported phylogenetic tree (with high bootstrap values). In practice, to find a such tree, the popular program RAxML \cite{Stamatakis21012014} is employed to compute the phylogenetic maximum-likelihood (ML) function with the following setup: the General Time Reversible model of nucleotide substitution with the $\Gamma$~model of rate heterogeneity and the hill-climbing optimization method, while the \textit{Prochlorococcus marinus} (NC\_009091.1) Cyanobacteria species is chosen as outgroup due to the supposed cyanobacteria origin of chloroplasts. The tree representation is obtained with Geneious~\cite{kearse2012geneious} based on the RAxML information. 

The first experiments are done using all five genes in the core genome of 98~species. Thus, in order to find a well supported phylogenetic tree from all core genes, which reflects a real evolutionary scenario, we have achieved high level calculations of bootstrapping values for 120~trees, by considering all permutations (using \emph{itertools} package) of gene orders\footnote{five core genes: $5!=120$ phylogenetic trees}. Among all these trees, we have then selected the tree with the largest value of its lowest bootstrap, this one is denoted as the most accurate tree~(MAT) in what follows,
after having verified that gene order has no effect on the supports. The MAT has a lowest bootstrap equal to 32 and to improve this value, we have investigated in a second stage of experiments whether some core genes are homoplasic ones. In fact, when the core is large enough, it is possible to remove a few of them that obviously break the supports according to the maximum likelihood inference. After having removing systematically 1, 2, 3, and 4~genes, the best phylogenetic tree, having its lowest bootstrap value equal to 35, was obtained after one~gene loss.

\begin{figure}[!ht]
\centering
    \includegraphics[width=0.9\textwidth]{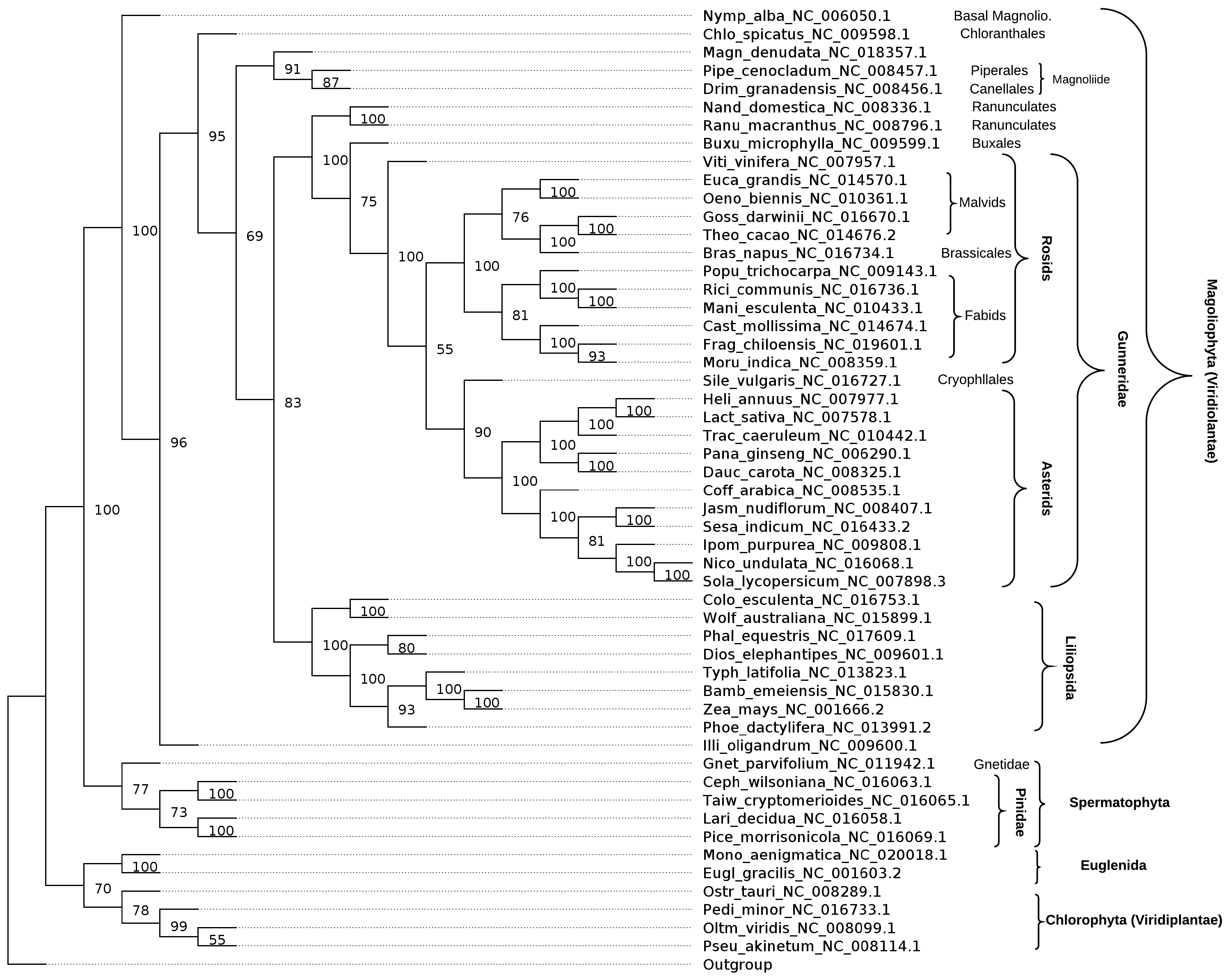}
    \caption{Core\_81 phylogenetic tree with 15 core genes (1 gene removed randomly).}\label{subfig-2:Phylo_55}
\end{figure}

The low improvement observed previously when removing some core genes suggests that their number is not sufficient to produce a well-supported phylogenetic tree. Therefore we decided for the next experiments to split the set of species in two and to work with the core genome of the largest subset: 52 genomes lead to a core genome of 16~genes\footnote{Core genes in Core\_81: \textit{psbE, psbD, petG, psbF, psbA, psbC, rpl36, psbN, psbI, psbJ, atpH, psaJ, atpI, atpA, psaA, and psaC.}} (Core\_81 in the core tree available online). As expected, working with this large core genome allows to really improve the lowest bootstrap value\footnote{The lowest bootstrap value for 16 core genes is 15.}, since by removing randomly 1, 2, 3, and 4 genes the resulting MAT has 55 for lowest bootstrap value. Figure~\ref{subfig-2:Phylo_55} presents this best tree obtained after removing one gene~(\emph{i.e.} \textit{atpI}). Let us notice that for large core genomes such an approach is intractable in practice, due to the dramatic number of core genes combinations to calculate.

\begin{figure}[!ht]
\centering
    \includegraphics[width=0.75\textwidth]{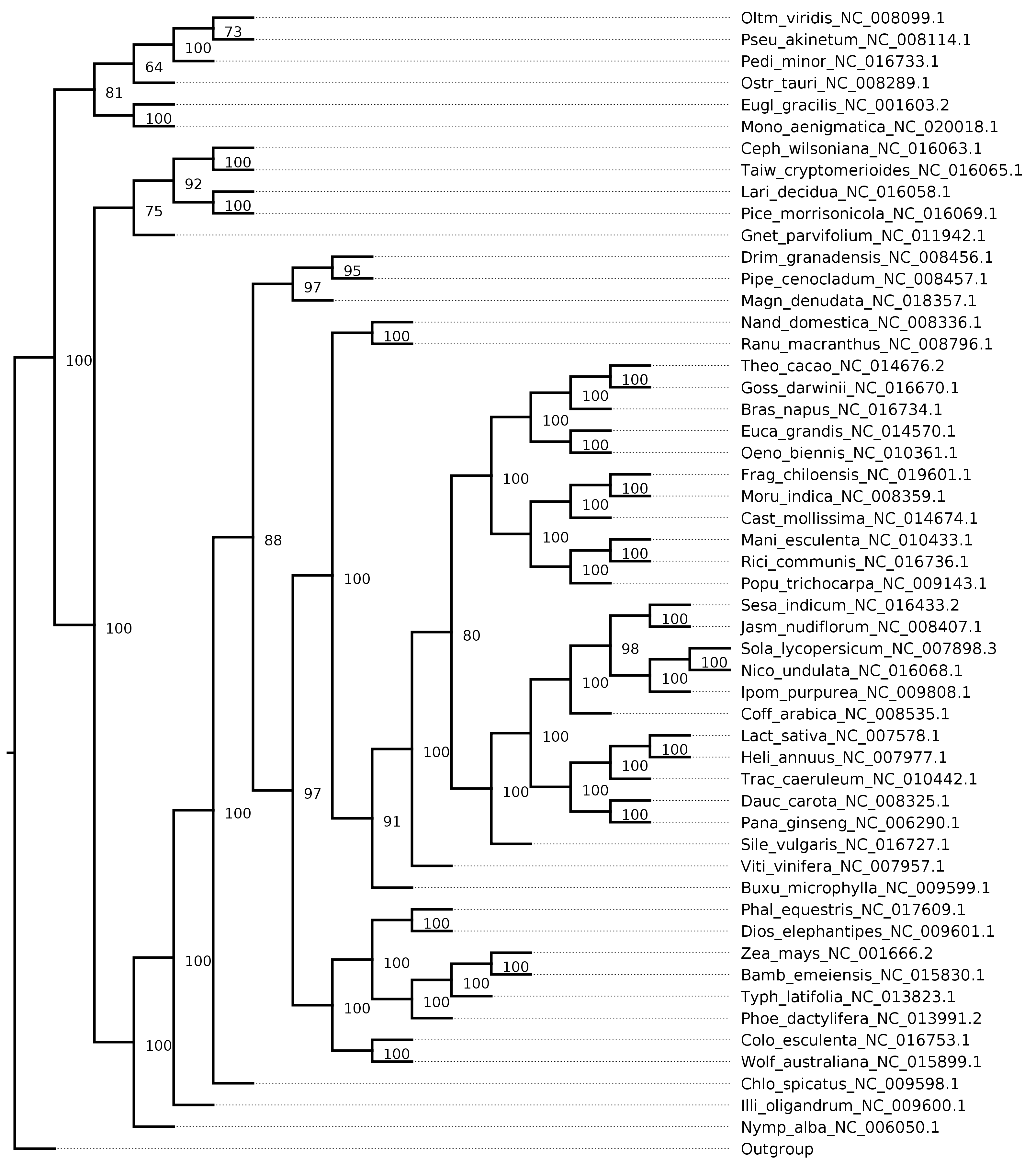}
    \caption{Supertree for Core\_81 from 248 bootstrap phylogenetic trees after removing 1, 2, 3, or 4 genes randomly.}\label{subfig-2:supertree}
\end{figure}

Finally, the support of the best phylogenetic tree can be improved again by using the whole knowledge inherited by all the previously constructed trees. \emph{SuperTripletes}~\cite{ranwez2010supertriplets} is one of the methods that can infer a supertree from a collection of bootstrapping phylogenetic trees. This tool\footnote{Available on~\url{http://www.supertriplets.univ-montp2.fr/index.php}} receives a file that stores all bootstrap values. In this last experiment, phylogenetic trees with 1, 2, 3, and 4 random gene loss have been concatenated in one file and transmitted to \emph{SuperTripletes}. The obtained supertree with all taxa is provided in Figure~\ref{subfig-2:supertree}. It can be seen that the minimum bootstrap has been further improved to 64.

\section{Conclusion}\label{sec:concl}
We have employed a ``quality test approach'' 
from previous study to extract core genes from a large set of chloroplastic genomes, and we 
compared it with the gene prediction approach developed also  
in our previous studies. 
A two stage similarity measure, on names and sequences, has thus been proposed for clustering DNA sequences in genes, 
which merges best results provided by NCBI and DOGMA. Results obtained
with this quality control test have finally been deeply compared with our previously obtained results.
Phylogenetic trees have finally been generated to investigate the distribution of chloroplasts and core genomes. High computations are made to produce the highest bootstrap values tree by generating all trees by considering gene orders and through removing randomly some genes from core genome. A supertree is then generated, showing a highly accurate phylogenetic tree for a large amount of plant species.


\bigskip
\textit{Computations have been performed on the supercomputer facilities of the M\'esocentre de calcul de Franche-Comt\'e.}

\bibliographystyle{unsrt}

\bibliography{biblio}


\end{document}